\documentclass[aps]{revtex4}

\usepackage{epsf}

\begin{document}

\title{Spontaneous decay of excited atomic states
near a carbon nanotube}

\author{I. V. Bondarev, G. Ya. Slepyan and S. A. Maksimenko}

\affiliation{The Institute for Nuclear Problems, The Belarusian
State University, Bobruiskaya Str.11, 220050 Minsk, BELARUS}

\begin{abstract}
Spontaneous decay process of an excited atom placed inside or
outside (near the surface) a carbon nanotube is analyzed.
Calculations have been performed for various achiral nanotubes.
The effect of the nanotube surface has been demonstrated to
dramatically increase the atomic spontaneous decay rate -- by 6 to
7 orders of magnitude compared with that of the same atom in
vacuum. Such an increase is associated with the nonradiative
decay via surface excitations in the nanotube.
\end{abstract}


\maketitle

%\newpage

Theoretical prediction of the Purcell effect in
1946~\cite{Purcell} has stimulated series of works, both
theoretical and experimental, aimed at the detailed investigation
of the phenomenon (see recent
papers~\cite{Dung,Sonder,Yang,Ando,Klimov} and references
therein). The effect lies in the fact that the spontaneous decay
rate of an excited atom essentially depends on whether or not the
atom is located near optical inhomogeneities and interfaces of
media with differing optical properties. Depending on the specific
configuration of inhomogeneities (interfaces), the atomic
spontaneous decay rate may both increase and decrease compared
with that of the same atom in free space. The Purcell effect took
on special significance recently in view of rapid progress in
physics of low-dimensional nanostructures. It was shown to be of
great importance for microcavities~\cite{Dung}, optical
fibers~\cite{Sonder}, photonic crystals~\cite{Yang},
semiconductor quantum dots~\cite{Sugawara}.

In Ref.~\cite{Klimov}, it was first suggested on the basis of the
model of an ideally conducting cylinder that spontaneous decay
process of excited atomic states near a carbon nanotube (CN)
might possess nontrivial peculiarities. However, the ideally
conducting cylinder is not quite an adequate model to describe the
optic properties of real CNs. It is our purpose in the present
paper to give consistent consideration to spontaneous decay
processes of the excited atom in the vicinity of CNs. We
calculate atomic spontaneous decay rate variation for infinitely
long achiral single-wall CNs of different radii and demonstrate
that the decay rate may dramatically increase due to the
nonradiative decay via CN surface excitations.

Quantum theory of the spontaneous decay of excited atomic states
in the vicinity of CN requires the solution of two fundamental
problems. They are (i) the problem of the macroscopic description
of the optical properties of solitary CN and (ii) the quantization
problem of an electromagnetic field in the presence of CN. We
solve problem (i) based upon the model described in
Refs.~\cite{Slepyan,Maks}. According to that model, CN is changed
by the infinitely thin anisotropically conducting cylinder with
effective boundary conditions imposed in such a way that the
field at a distance from the cylinder surface be identical to the
actual electromagnetic field excited in the system. In so doing,
only the axial conductivity of CN is taken into account and the
transverse conductivity is neglected for the following reason.
The axial conductivity forms by intraband and direct interband
transitions of $\pi$-electrons in CN, while the transverse one
only forms by indirect interband transitions~\cite{Tasaki,Saito}.
The intraband transitions dominate at lower frequencies and the
interband ones start essentially contributing to the total CN
conductivity at higher frequencies. However, the contribution of
the indirect interband transitions, being strongly suppressed by
depolarization fields, is always smaller compared with that of
the direct interband transitions~\cite{Tasaki,Saito}. We use the
dispersion law for $\pi$-electrons in the tight-binding
approximation with allowance made for the azimuthal momentum
quantization. Energy dissipation is taken into account within the
relaxation time approximation. The spatial dispersion of the CN
conductivity is neglected (see Refs.~\cite{Slepyan,Maks} for its
role in CNs).

The quantization procedure of the electromagnetic field,
problem~(ii), faces difficulties similar to those in quantum
optics of 3D Kramers-Kronig media where the canonical quantization
scheme commonly used does not work since, because of absorption,
the respective operator Maxwell equations become non-Hermitian.
A~standard approach overcoming these difficulties involves a noise
current term incorporated into the operator Maxwell
equations~\cite{Vogel}. We use the analogous approach to quantize
the electromagnetic field in the presence of CN. In this case, the
noise current becomes the surface one and, therefore, may be
incorporated into the boundary conditions for the Maxwell
equations rather than into the Maxwell equations themselves. As
this takes place, the effective boundary conditions for electric
field and magnetic field operators in frequency-domain space take
the form
\[
\mathbf{n}\times\left(\left.\hat{\mathbf{E}}_{_{_{_{_{_{}}}}}}\!\right|_{r=R_{cn}+0}
\!\!-\!\left.\hat{\mathbf{E}}_{_{_{_{_{_{}}}}}}\!\right|_{r=R_{cn}-0}\right)=0\,,
\]\vspace{-0.9cm}
\begin{equation}
\label{boundary}
\end{equation}\vspace{-0.9cm}
\[
\mathbf{n}\times\left(\left.
\hat{\mathbf{H}}_{_{_{_{_{_{}}}}}}\!\right|_{r=R_{cn}+0}\!\!-\!\left.
\hat{\mathbf{H}}_{_{_{_{_{_{}}}}}}\!\right|_{r=R_{cn}-0}\right)
+{4\pi\over{c}}\,\hat{J}_{z}^{N}\,\mathbf{e}_{z}=
{4\pi\over{c}}\,\sigma_{zz}(\omega)\hat{E}_{z}\,\mathbf{e}_{z}\;,
\]
where $r$ is the radial spatial coordinate, $R_{cn}$ the CN
radius, $\mathbf{n}$ and $\mathbf{e}_{z}$ are the unit vectors
along the external normal to the CN surface and along the CN axis,
respectively, $\sigma_{zz}(\omega)$ is the axial dynamical
conductivity of CN (see Eq.(36) in Ref.~\cite{Slepyan}),
$\hat{J}_{z}^{N}$ is the axial noise current operator. The latter
one is expressed in terms of 2D scalar bosonic field operator
$\hat{f}(\mathbf{R})$ ($\mathbf{R}\!\in\!$ CN surface) as
$\hat{J}_{z}^{N}\!=\!\sqrt{\hbar\omega\mbox{Re}\,\sigma_{zz}(\omega)/\pi}\,
\hat{f}(\mathbf{R})$ and is responsible for correct commutation
relations of $\hat{\mathbf{E}}$ and $\hat{\mathbf{H}}$
operators~\cite{Vogel}. Homogeneous Maxwell equations along with
boundary conditions~(\ref{boundary}) describe quantum
electrodynamics of CNs.

Let the excited atom with an electric dipole transition allowed be
located in the vicinity of CN and let the atomic dipole moment be
oriented along the CN axis. Following the quantization scheme
above, one can by analogy with Ref.~\cite{Dung} obtain the
Volterra integral equation for atomic decay dynamics from the
upper stationary state to the lower one
\begin{equation}
C_{u}(t)=1+\int_{0}^{t}\!K(t-t{^\prime})\,C_{u}(t^{\prime})\,dt^{\prime}
\label{Volterra}
\end{equation}
with $C_{u}$ being the occupation probability amplitude of the
upper state and the kernel given by
\begin{equation}
K(\tau)={4\,k_{A}^{2}|\,\mu_{z}|^{2}\over{\hbar}}\int_{0}^{\infty}
\!\!d\omega\,{\mbox{Im}\,G_{zz}(\mathbf{r}_{A},\mathbf{r}_{A},\omega)
\over{i\,(\omega-\omega_{A})}}\left[\,e^{\textstyle-i(\omega-\omega_{A})\tau}
-1\right], \label{kernel}
\end{equation}
where $\mathbf{\mu}_{z}$ and $\omega _{A}$ are the matrix element
and the frequency of the atomic dipole transition, respectively,
$\mathbf{r}_{A}$ is the radius-vector of the atomic position,
$k_{A}=\omega _{A}/c$, and
$G_{zz}(\mathbf{r},\mathbf{r}_{A},\omega)$ is the axial component
of the classical electromagnetic field Green tensor in the
presence of~CN.

For further calculations we need Green tensor components
$G_{\alpha z}$ ($\alpha\!=\!r,\varphi,z$ are the cylindrical
coordinates associated with CN). We use the representation
$G_{\alpha z}\!=\!(k^{-2}\partial_{\alpha}\partial_{z}+
\delta_{\alpha z})\,g$ with $g(\mathbf{r},\mathbf{r}_{A},\omega)$
being the scalar Green function of the electromagnetic field in
the presence of CN. The latter one is expanded over cylindrical
waves in terms of boundary conditions~(\ref{boundary}), so that
at $r,r_{A}\!>\!R_{cn}$ one obtains
\begin{equation}
g=g_{0}-{R_{cn}\over{(2\pi)^{2}}}\sum_{p=-\infty}^{\infty}\!\!
e^{ip\varphi}\!\!\int_{C}{\beta(\omega)\,v^{2}I_{p}^{2}(vR_{cn})
K_{p}(vr_{A})K_{p}(vr)\over{1+\beta(\omega)\,v^{2}R_{cn}
I_{p}(vR_{cn})K_{p}(vR_{cn})}}\;e^{\textstyle ihz}dh\,, \label{g}
\end{equation}
where $g_{0}\!=\exp(ik\,|\mathbf{r}-\mathbf{r}_{A}|)/4\pi
|\mathbf{r}-\mathbf{r}_{A}|$ is the free space Green function,
$\beta(\omega)=4\pi i\,\sigma_{zz}(\omega)/\omega$,
$v\!=\!\sqrt{h^{2}-k^{2}}$, $I_{p}(X)$ and $K_{p}(X)$ are the
modified cylindrical Bessel functions. Integration contour $C$
goes along the real axis of the complex plane and envelopes
branch points $\pm k\,$ from below and from above, respectively.
For $r,r_{A}\!<\!R_{cn}$, Eq.(\ref{g}) is modified by the
replacement $I_{p}\leftrightarrow K_{p}$ in the numerator of the
integrand.

Consider decay process in the Markovian approximation. Then,
factor $[\,\exp(-ix\tau)-\!1]/ix$ in Eq.(\ref{kernel}) is changed
by $\pi\delta(x)+i\,{\cal{P}}(1/x)$ (${\cal{P}}$ denotes the
principal value), and one arrives at the exponential decay model
with $K(\tau)\!=\!-\Gamma/2+i\,\delta\omega$, where $\Gamma$ and
$\delta\omega$ are the decay rate and the Lamb shift of the upper
atomic level, respectively. For the outward atomic position
($r_{A}\!>\!R_{cn}$), the decay rate is written in view of
Eqs.(\ref{Volterra})-(\ref{g}) as
\begin{equation}
{\Gamma\over{\Gamma_{0}}}=\xi(\omega_{A})=1+{3R_{cn}\over{2\pi
k_{A}^{3}}}\sum_{p=-\infty}^{\infty}\int_{C}{\beta_{A}\,v_{A}^{4}
I_{p}^{2}(v_{A}R_{cn})K_{p}^{2}(v_{A}r_{A})
\over{1+R_{cn}\,\beta_{A}\,v_{A}^{2}
I_{p}(v_{A}R_{cn})K_{p}(v_{A}R_{cn})}}\;dh\,, \label{ksi}
\end{equation}
with $\beta_{A}\!=\!\beta(\omega_{A})\,$,
$v_{A}\!=\!\sqrt{h^{2}-k_{A}^{2}}\,$,
$\,\Gamma_{0}\!=4k_{A}^{\,3}\,|\,\mu_{z}|^{2}/3\hbar\,$ being the
free space decay rate and $\xi(\omega_{A})$ representing the
influence of CN. For the inward position ($r_{A}\!<\!R_{cn}$),
Eq.(\ref{ksi}) is modified by the simple replacement
$r_{A}\leftrightarrow R_{cn}$ in the numerator of the integrand.
Note the divergence of the integral in Eq.(\ref{ksi}) at
$r_{A}\!=\!R_{cn}\,$, i.~e. when the atom is located directly on
the CN surface. This divergence originates from the averaging
procedure over physically infinitely small volume when describing
CN optical properties. Such an averaging does not assume any
additional atoms on the CN surface, to take them into
consideration the procedure must be modified. Thus, the
applicability domain of our model is restricted by the condition
$\mid\!r_{A}\!-R_{cn}\!\mid\,>\!a$, where $a=1.42$~\AA\space is
the interatomic distance in CN.

The decay of the excited atom interacting with medium may proceed
both via real photon emission (radiative decay) and via virtual
photon emission with subsequent medium quasiparticle excitation
(nonradiative decay). Both of these decay channels are present in
atomic spontaneous decay rate $\Gamma$ described by
Eq.(\ref{ksi}). The problem of the total $\Gamma$ partition to
radiative and nonradiative contributions is not trivial.
Radiative contribution $\Gamma_{r}$ was estimated by using a
Poynting vector approach for the atom near a microsphere
in~\cite{Dung} and for the atom inside an optic fiber
in~\cite{Sonder}. Following this approach, we estimate the
spontaneous radiation intensity distribution
$I(\mathbf{r},t)\!=\,\mid\!\mathbf{F}(\mathbf{r},\mathbf{r}_{A},\omega_{A})
\!\mid^{2}\!\exp(-\Gamma t)$, where
\[
F_{\alpha}(\mathbf{r},\mathbf{r}_{A},\omega_{A})=-4\pi
i\,k_{A}^{2}\,\mu_{z}\left\{G_{\alpha
z}(\mathbf{r},\mathbf{r}_{A},\omega_{A})-{1\over{\pi}}\,{\cal{P}}\!\!
\int_{0}^{\infty}\!{\mbox{Im}\,G_{\alpha
z}(\mathbf{r},\mathbf{r}_{A},\omega)\over{\omega
+\omega_{A}}}\,d\omega\right\},
\]
at large distances $|\mathbf{r}|\rightarrow\infty$. In so doing,
the second term does not contribute and the contribution of the
first term is easily found by the stationary phase
method~\cite{Dingle} to give in the spherical coordinates
($|\mathbf{r}|$, $\phi$, $\theta$) associated with the atom
\[
\lim_{|\mathbf{r}|\rightarrow\infty}\mathbf{F}(\mathbf{r},\mathbf{r}_{A},
\omega_{A})\simeq-ik_{A}^{2}\,\mu_{z}{e^{\textstyle ik_{A}|\mathbf{r}|}
\over{|\mathbf{r}|}}\;\mathbf{e}_{\theta}\sin\theta\!\!
\sum_{p=-\infty}^{\infty}\!\!\Xi_{p}(-ik_{A}\sin\theta)\,
e^{\textstyle ip\,\phi}
\]
with $\mathbf{e}_{\theta}$ being the spherical ort and
\[
\Xi_{p}(X)=\left\{
\begin{array}{cc}{\displaystyle I_{p_{_{}}}(Xr_{A})
\over{\displaystyle1+R_{cn}\,\beta_{A}X^{2^{^{^{}}}}\!
I_{p}(XR_{cn})K_{p}(XR_{cn})}}\;,&\mbox{if~~}r_{A}\!<\!R_{cn}\\[0.5cm]
I_{p}(Xr_{\!_{A}})-{\displaystyle~R_{cn}\,\beta_{A}X^{2^{^{^{}}}}\!
I_{p_{_{}}}^{2}(XR_{cn})K_{p}(Xr_{A})
\over{\displaystyle1+R_{cn}\,\beta_{A}X^{2^{^{^{}}}}\!
I_{p}(XR_{cn})K_{p}(XR_{cn})}}\;,&\mbox{if~~}r_{A}\!>\!R_{cn}\,.\end{array}\right.
\]
The relative contribution of the radiative decay is then given by
\[
{\Gamma_{r}\over{\Gamma}}={c\over{2\pi\hbar\omega_{A}}}
\lim_{|\mathbf{r}|\rightarrow\infty}\int_{0}^{\infty}\!\!\!dt\!
\int_{0}^{2\pi}\!\!\!d\phi\!\int_{0}^{\pi}\!\!|\mathbf{r}|^{2}\,
I(\mathbf{r},t)\,\sin\vartheta\,d\vartheta
\]\vspace{-1.1cm}
\begin{equation}
\label{rad}
\end{equation}\vspace{-1.1cm}
\[\hskip0.6cm
={3\over{4\,\xi(\omega_{A})}}\sum_{p=-\infty}^{\infty}\int_{0}^{\pi}\!\!
\mid\!\Xi_{p}(-i k_{\!_{A}}\!\sin\vartheta)\!\mid^{2}\!\sin^{3}\!
\vartheta\,d\vartheta\;.
\]

\begin{figure}[p]
\epsfxsize=16cm \centering{\epsfbox{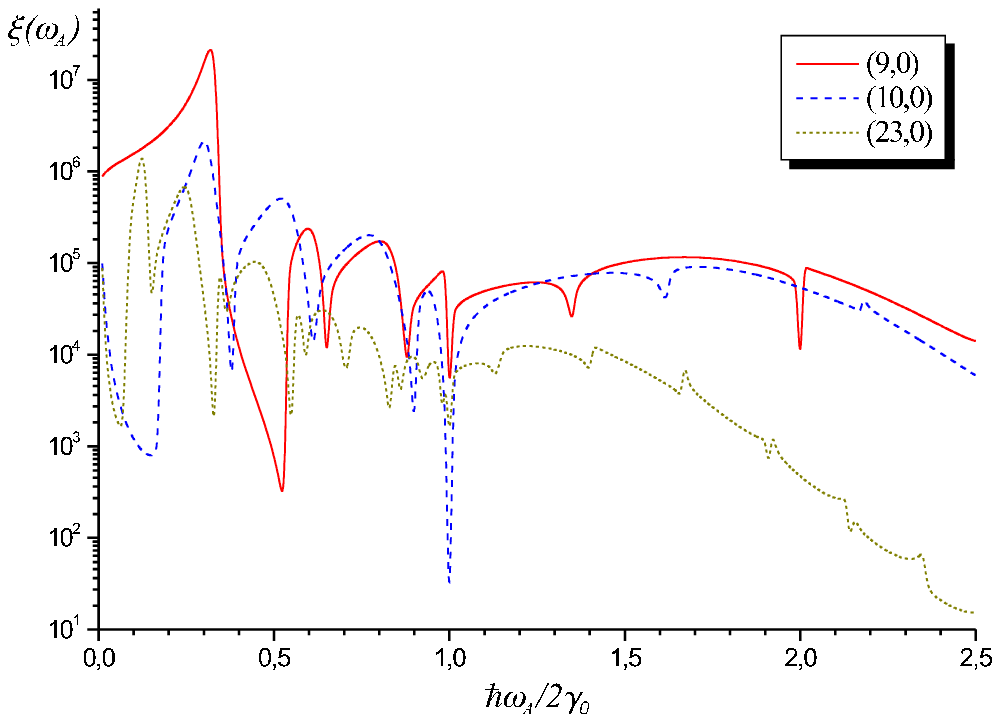}}
\caption{}
\label{fig1}
\end{figure}

Figure~\ref{fig1} shows the results of the numerical calculations
of factor $\xi(\omega_{A})$ according to Eq.(\ref{ksi}) for
metallic and semiconducting CNs of $(m,0)$ type ("zigzag"). The
atom is located on the symmetry axis inside CN. The presence of
CN is seen to drastically accelerate spontaneous decay process of
an excited atomic state. Frequency range
$0.305\!<\!\hbar\omega_{A}/2\gamma_{0}\!<\!0.574$
($\gamma_{0}\!=\!2.7$~eV is the carbon overlap integral)
corresponds to visible light. Lower frequencies
$\hbar\omega_{A}/2\gamma_{0}\!<\!0.305$ correspond to infra-red
waves emitted by highly excited Rydberg atomic states. At these
frequencies, the large difference ($3\!-\!4$ orders of magnitude)
is seen in the value of $\xi(\omega_{A})$ for metallic
($m\!=\!3q$, $q\!=\!1,2,...$) and semiconducting ($m\!\ne\!3q$)
CNs. The difference is caused by the Drude-type conductivity
(intraband electronic transitions) dominating in this region,
whose relative contribution to the total CN conductivity is
larger in metallic CNs than in semiconducting
ones~\cite{Slepyan,Maks,Tasaki,Saito}. As the frequency
increases, interband electronic transitions start manifesting
themselves and function $\xi(\omega_{A})$ becomes irregular. At
high frequencies, there is no significant difference between
metallic and semiconducting CNs of close radii. Function
$\xi(\omega_{A})$ has dips when $\omega_{A}$ equals the interband
transition frequencies; in particular, there is a dip at
$\hbar\omega_{A}\!=\!2\gamma_{0}$ for all CNs considered. It is
essential that $\xi(\omega_{A})\!\gg\!1$ throughout the entire
frequency range considered. This lets us formulate the central
result of the present paper: {\it the spontaneous decay
probability of the atom in the vicinity of CN is larger by a few
orders of magnitude than that of the same atom in free space}. In
other words, the Purcell effect is extraordinarily strong in CNs.
This is physically explained by the photon vacuum
renormalization: the density of photonic states (and, as a
consequence, the atomic decay rate) near CN effectively increases
($\rho^{\mathit~\!\!ef\!f}\!(\omega)\,d\omega\!=
\xi(\omega)\,\omega^{2}d\omega/\pi c^{3}$) since, along with
ordinary free photons, there appear the photonic states coupled
with CN electronic quasiparticle excitations.

In Refs.~\cite{Slepyan,Maks}, the possibility was shown of the
existence of slow surface electromagnetic waves in CNs. In
Ref.~\cite{Dung}, such waves were shown to be responsible for the
strong Purcell effect for the atom in the spherical microcavity.
The results of the present paper agree qualitatively with those
obtained in Ref.~\cite{Dung}. Quantitatively,
$\,\xi_{max}\!\sim\!10^{6}-10^{7}$ for CNs while
$\xi_{max}\!\sim\!10^{4}$ for the microcavity~\cite{Dung}, i.~e.
the Purcell effect is much stronger in CNs. It is worth noting
that for the atoms with large enough $\Gamma_{0}$, there is the
risk of going beyond the applicability limits of the two-level
model and Markovian approximation. Then, in the first case, the
mutual overlap is possible of the levels due to their strong
broadening and, in the second one, the atomic spontaneous decay
may become nonexponential so that the problem may require the
numerical solution of integral equation (\ref{Volterra}) with
kernel (\ref{kernel}).

\begin{figure}[p]
\epsfxsize=16cm \centering{\epsfbox{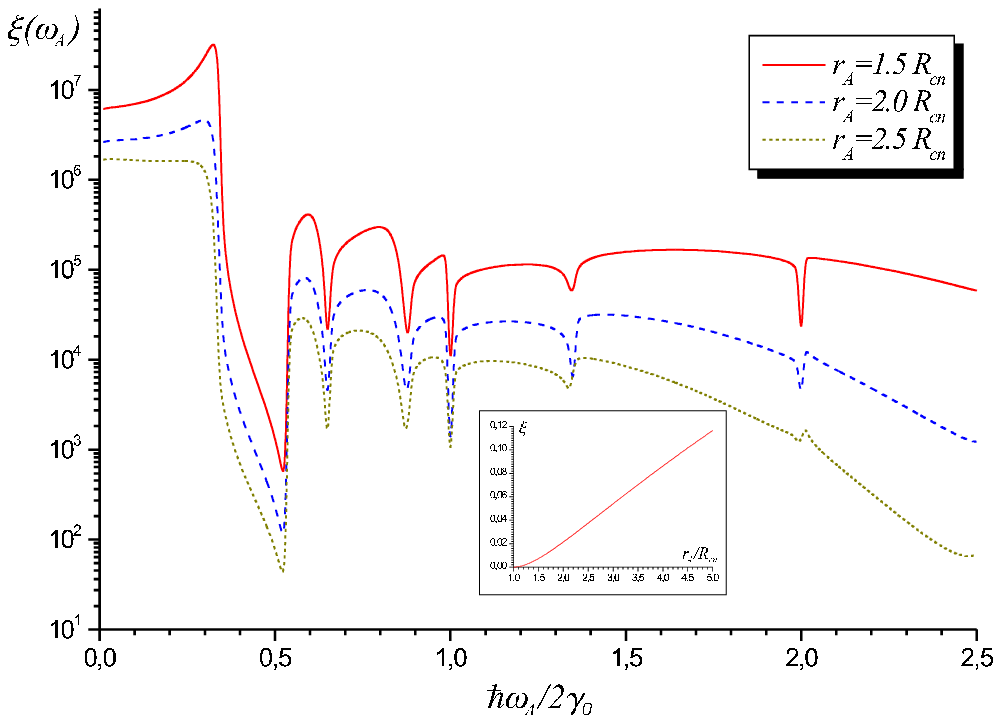}}
\caption{}
\label{fig2}
\end{figure}

Figure~\ref{fig2} shows $\xi(\omega_{A})$ for the atom located
outside CN at different distances from the CN surface. The
qualitative behavior of $\xi(\omega_{A})$ is similar to that
represented in Figure~1 for the atom inside CN. Factor
$\xi(\omega_{A})$ is seen to rapidly decrease with raising the
distance as it should be in view of the evident fact that
photonic states coupled with CN electronic excitations (those
increasing effective density $\rho^{\mathit~\!\!ef\!f}\!(\omega)$)
are spatially localized on the CN surface and their coupling
strength with the excited atom decreases with raising the distance
of the atom from~CN.

\begin{figure}[p]
\epsfxsize=16cm \centering{\epsfbox{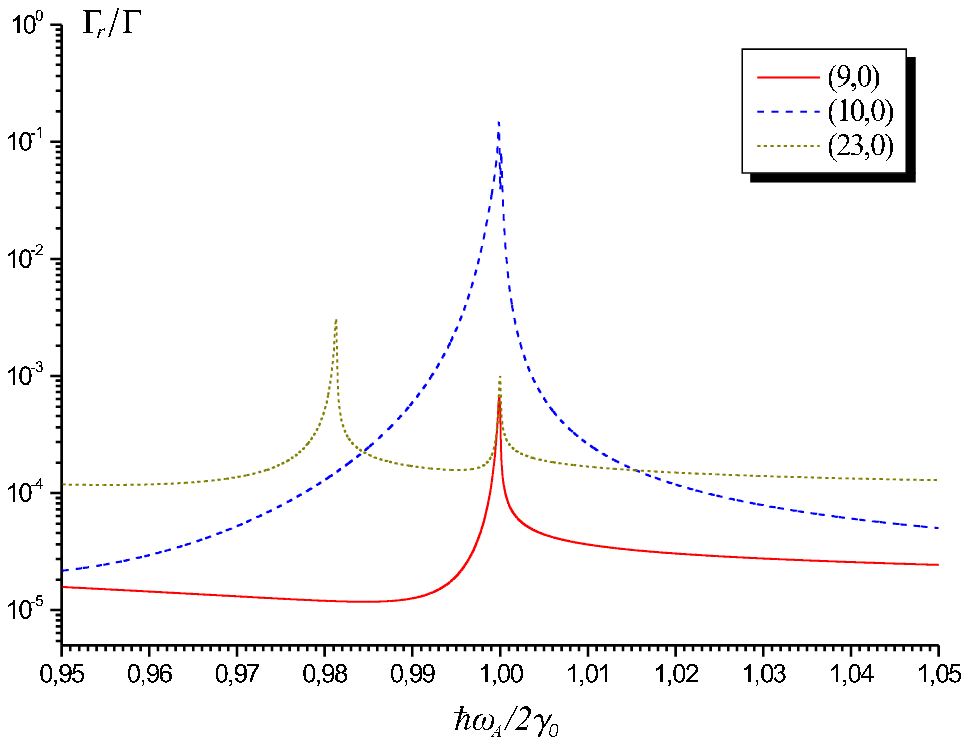}}
\caption{}
\label{fig3}
\end{figure}

Figure~\ref{fig3} shows the ratio $\Gamma_{r}/\Gamma$ calculated
according to Eq.(\ref{rad}) for the atom in the centre of
different CNs. (Note that
$\Gamma_{r}/\Gamma\!=\!W_{s}(\omega_{A})/\hbar\omega_{A}$ with
$W_{s}(\omega_{A})$ being the total power of the atomic
spontaneous radiation far from CN.) The ratio is very small
indicating that the nonradiative decay dominates. However, the
radiative decay is seen to essentially contribute in the vicinity
of the interband transition frequencies so that the frequency
dependence of $W_{s}(\omega_{A})$ (which, in principle, can be
measured experimentally) reproduces CN electronic structure
peculiarities. The main conclusion one can draw from Figure~3 is
{\it the Purcell effect in CNs, along with the increase of the
atomic spontaneous decay rate, manifests itself by decreasing the
power of the spontaneous radiation}.

Our model of the atomic spontaneous decay in the presence of CN
includes, as a limiting case, the model of the ideally conducting
cylinder considered in Ref.~\cite{Klimov}. In particular, our
Eq.(\ref{ksi}) reduces for the outward atomic position to
Eqs.(15),(18) of Ref.~\cite{Klimov} as
$\sigma_{zz}\rightarrow\infty$. The inset in Figure~\ref{fig2}
shows factor $\xi(\omega_{A})$ at
$\omega_{A}\!=3\gamma_{0}/\hbar$ ($k_{A}R_{cn}\!\simeq0.01$) as a
function of $r_{A}/R_{cn}$ for this case. The dependence is
similar to that reported in Ref.~\cite{Klimov} for z-oriented
dipole at $k_{A}R_{cn}\!=1$. For the atom inside CN,
Eq.(\ref{ksi}) yields $\xi(\omega_{A})\rightarrow 0$ as
$\sigma_{zz}\rightarrow\infty$
--- the result is natural since in this case only one eigen
electromagnetic mode can propagate in CN; this mode is
essentially transverse and, consequently, is not coupled with the
atomic dipole moment oriented longitudinally. However, the actual
$\xi(\omega_{A})$ behavior discussed above is quite different
from that predicted by the ideally conducting cylinder model
since the latter one does not account for CN electronic
quasiparticle excitations responsible for the nonradiative atomic
decay dominating the total spontaneous decay process.

Our theory may be generalized to cover the transverse atomic
electric dipole orientation, electric quadrupole and magnetic
dipole atomic transitions, properties of organic molecules
inside/outside CNs~\cite{Petrov} and of fullerene
peapods~\cite{Liu}. The mechanism revealed of the photon vacuum
renormalization is likely to manifest itself in other phenomena
in CNs such as, for example, Casimir forces, electromagnetic
fluctuations, etc.~\cite{Bordag}.

The results of the present work may be tested by methods of
atomic fluorescent spectroscopy and may possess various physical
consequences. In particular, the effect of the drastic increase of
the atomic spontaneous decay rate may turn out to be of practical
importance in problems of the laser control of atomic
motion~\cite{Scully}, yielding the increase of the ponderomotive
force acting on the atom moving in the vicinity of CN in a laser
field. One might expect the Purcell effect peculiarities that we
predict for CNs to manifest in macroscopic anisotropically
conducting waveguides with strong wave deceleration (for example,
in microwave spiral or collar waveguides with highly excited
Rydberg atoms inside).

The work was financially supported by the Foundation for Basic
Research of the National Academy of Sciences of the Republic of
Belarus (Grants No~F01-176 and No~F02R-047).

\newpage

\noindent{\bf Figure captions:}

\vskip0.5cm

\noindent{\bf Figure~\ref{fig1}:} Factor $\xi(\omega_{A})$
calculated from Eq.(\ref{ksi}) for the atom in the center of
different "zigzag" CNs. Surface axial conductivities
$\sigma_{\!zz}$ appearing in Eq.(\ref{ksi}) were calculated in the
$\tau$-approximation with $\tau=3\times10^{-12}$~s$^{-1}$.

\vskip0.5cm

\noindent{\bf Figure~\ref{fig2}:} Factor $\xi(\omega_{\!_{A}})$
for the atom at different distances outside "zigzag" (9,0) CN.
Inset: $\xi(\omega_{A})$ at $\omega_{A}\!=\!3\gamma_{0}/\hbar$ as
a function of $r_{A}/R_{cn}$ for the atom near (9,0) CN in the
model of the ideally conducting cylinder.

\vskip0.5cm

\noindent{\bf Figure~\ref{fig3}:} Ratio $\Gamma_{r}/\Gamma$
calculated from Eq.(\ref{rad}) for the atom in the center of
different "zigzag" CNs.

\newpage

\end{document}